\documentclass[reprint,aps,pra,amsmath,amssymb,showpacs,superscriptaddress,]{revtex4-1}
\usepackage{graphicx}
\usepackage{dcolumn}
\usepackage{bm}
\usepackage{color}
\usepackage{upgreek}

\begin{document}

\title{Alignment dependent ultrafast electron-nuclear dynamics in high-order harmonic generation}

\author{Mu-Zi Li}\affiliation{State Key Laboratory of Magnetic Resonance and Atomic and Molecular Physics, Wuhan Institute of Physics and Mathematics, Chinese Academy of Sciences, Wuhan 430071, China}\affiliation{University of Chinese Academy of Sciences, Beijing 100049, China}

\author{Guang-Rui Jia}\affiliation{State Key Laboratory of Magnetic Resonance and Atomic and Molecular Physics, Wuhan Institute of Physics and Mathematics, Chinese Academy of Sciences, Wuhan 430071, China}\affiliation{College of Physics and Materials Science, Henan Normal University, Xinxiang 453007, China}

\author{Xue-Bin Bian}\email{xuebin.bian@wipm.ac.cn}\affiliation{State Key Laboratory of Magnetic Resonance and Atomic and Molecular Physics, Wuhan Institute of Physics and Mathematics, Chinese Academy of Sciences, Wuhan 430071, China}

\begin{abstract}

We investigated the high-order harmonic generation (HHG) process of diatomic molecular ion $\mathrm{H}_2^+$ in non-Born-Oppenheimer approximations. The corresponding three-dimensional time-dependent Schr\"odinger equation is solved with arbitrary alignment angles. It is found that the nuclear motion can lead to spectral modulation of HHG. Redshifts are unique in molecular HHG which decrease with the increase of alignment angles of the molecules and are sensitive to the initial vibrational states. It can be used to extract the ultrafast electron-nuclear dynamics and image molecular structure.


\end{abstract}

\maketitle

\section{INTRODUCTION}\label{I}

When matters are subjected to intense laser fields, high-order harmonic generation (HHG) may occur \cite{McPherson, Huillier, Protopapas}. In the semiclassical three-step model, free electrons are considered to do classical motions driven by the laser field after they are tunneling ionized from the ground state. Then a part of the free electrons return to the vicinity of the parent ions and HHGs yield by the recombination of them \cite{Corkum}. Explaining the cutoff and plateau of HHG excellently, this model is widely accepted.

The dynamics in molecules are much more complicated than that in atoms for the extra nuclear motions such as vibration and rotation. Therefore many new phenomena occur in the ultrafast processes within molecules. There are two or more nuclei in molecules, which implies more recollision channels. Some effects like charge-resonance enhanced ionization (CREI) and transient electron localization can enhance the ionization process strongly \cite{Zuo, Takemoto}. In the CREI, a pair of charge-resonant states strongly couple at a critical range of internuclear separation, which leads to a much higher ionization rate \cite{Zuo, Constant, Chelkowski, Itzhak}. These effects are verified to remain robust in non-Born-Oppenheimer approximations (NBOA) \cite{Takemoto}. Besides, there are bond softening \cite{Bucksbaum}, bond hardening \cite{Bandrauk, Zavriyev}, and above-threshold dissociation \cite{Suzor}, etc. The HHG process is strongly related to the ionization of molecules. The nuclear motion modifies electron dynamics and influences the ionization processes dramatically \cite{Khosravi}, thus it's crucial to study HHGs in NBOA. Considering these effects, HHGs can be applied not only in the generation of attosecond laser pulses \cite{Corkum0}, but also in resolving dynamics on their natural attosecond time scale.

The information of nuclear dynamics can be extracted from frequency modulation (FM) \cite{Bian} as well as amplitude modulation (AM) of HHG spectra \cite{Lein, Baker}. The former is due to inter-cycle electron dynamics, while the latter comes from intra-cycle dynamics. However, AM may be easily influenced by the multichannel interference effects and  the intrinsic energy-dependent ionization or rescattering cross sections. FM is more stable to probe the nuclear motion. Several phenomena such as depletion and propagation effects can lead to FM of HHG spectra \cite{Brandi, Shin}. However, both of them only lead to blueshifts in atomic HHG. Redshifts are found to be unique in molecular HHG spectra theoretically \cite{Bian} and experimentally \cite{Yuan, He} due to the nuclear motion, which is insensitive to interference effects and rescattering sections. The increase or decrease of the effective amplitude of the laser fields in different optical cycles causes the phase \cite{Kan} and therefore the frequency of each harmonics changing \cite{Chang}. The nuclear motions enlarge the above effect and are recorded in the modulation of HHG frequencies \cite{Bian, Du, Ahmadi}. Based on this, several works about the detection of molecular motions by the FM of HHG spectra have been done theoretically \cite{Bian, Ahmadi} and experimentally \cite{Yuan, He}. However, all these studies have only considered the situation that the molecules are initially on the ground state and the molecular axis is parallel to the laser polarization direction. In fact, the linear molecules may be excited and randomly aligned. In this work, we study the influences of the initial vibrational states and the alignment dependence of diatomic molecules on redshifts of HHGs in NBOA by numerically calculating the interaction of three-dimensional $\mathrm{H}_2^+$ molecules in linearly polarized pulses, which to our knowledge has not been investigated yet. Our results show that they play crucial roles. Thus the nuclear motion information is expected to be extracted from the FM of HHG spectra.
\section{NUMERICAL METHODS}\label{II}
\begin{figure*}
\centering\includegraphics*[width=\textwidth]{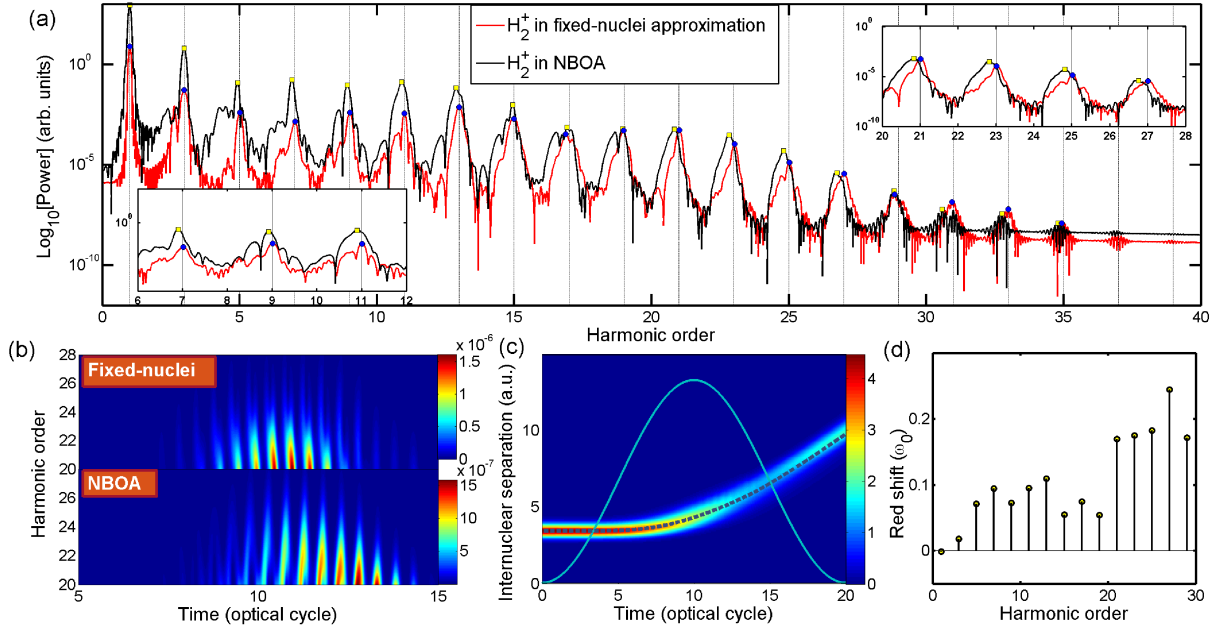}
\caption{(Color online) HHG of $\mathrm{H}_2^+$ in NBOA and fixed-nuclei approximations with molecular axis parallel to the laser polarization direction. The laser wavelength is 500nm, and the intensity is $10^{14}\mathrm{W}/\mathrm{cm}^2$. In (a), redshifts occur in NBOA (black line) compared with the fixed-nuclei approximation (red line) where the internuclear separation is fixed at 3 a.u. The time profiles in (b) show the time of the generation of harmonics in high energy region in NBOA (lower panel) and in the fixed-nuclei approximations (upper panel). (c) shows the time evolution of the nuclear probability density and the time-dependent internuclear separation (gray dotted line). For illustration, the laser pulse envelope multiplied by 13 is plotted (cyan line). In (d), redshifts tend to increase with the increase of harmonic orders.}
 \label{Fig1}
\end{figure*}

In order to investigate the alignment dependence, we use a 3D time-dependent Schr\"odinger equation (TDSE)  in the velocity gauge with an additional nuclear vibration freedom  to describe the interaction. We solve it in the molecular frame coordinates. The electron motion is restricted in a 2D plane and the nuclear vibration is along the molecular axis (atomic units are used) \cite{Chelkowski0}.
\begin{equation}\label{E1}
i \frac{\partial \psi(x, y, R, t)}{\partial t} = H(x, y, R, t) \psi(x, y, R, t),
\end{equation}
where
\begin{equation}\label{E2}
H(x, y, R, t) = H_{e}(x, y, t) + V(x, y, R) + H_{N}(R),
\end{equation}
\begin{equation}\label{E3}
H_{e}(x, y, t) = - \beta \left(\frac{\partial^2}{\partial x^2} + \frac{\partial^2}{\partial y^2}\right) - \kappa \textbf{p}_{e} \cdot \textbf{A}(t),
\end{equation}
\begin{equation}\label{E4}
H_{N}(R) = - \frac{1}{m_p} \frac{\partial^2}{\partial R^2} + \frac{1}{R},
\end{equation}
\begin{equation}\label{E5}
V(x, y) = - \sum_{k=1,2} \frac{1}{\sqrt{(x-x_k)^2 + (y-y_k)^2 + 1}},
\end{equation}
\begin{equation}\label{E6}
\beta = \frac{2 m_p + m_e}{4 m_p m_e}, \kappa = 1 + \frac{m_e}{2 m_p + m_e},
\end{equation}

$\textbf{A}(t) = - \int_0^t \textbf{E}(t') \mathrm{d}t'$, is the vector potential of the external field. $m_e$ and $m_p$ are the masses of the electron and the proton ($m_e=1, m_p=1837$) respectively. $\textbf{E}(t) = \left(E(t) \cos \theta, E(t) \sin \theta)\right)$, where $\theta$ is the angle between the molecular axis and the laser polarization direction and $E(t) = E_0 f(t) \cos (\omega_0 t)$. $f(t)$  is the envelope of the pulse which is a cosine-square shape with total duration of 20 cycles. $x_{1/2} = \pm \textbf{R} / 2$ and $y_{1/2} = 0$. In the fixed-nuclei approximation, $\textbf{R}$ as the internuclear separation is fixed at 3 a.u. However in NBOA, $\textbf{R}$ is also a variable. We solve the TDSE by using the second-order split-operator method, which can preserve unitarity in the procedure of the iteration \cite{Bandrauk0, Feit}. A 500 nm, $10^{14}\mathrm{W}/\mathrm{cm}^2$ linearly polarized pulse is used. We calculate 2014 steps for each cycle. The TDSE is solved for $|x| < 55$ a.u., $|y| < 55$ a.u. with $dx = dy = 0.16$ a.u., and $0 < R < 30$ a.u. with $dR = 0.083$ a.u. We apply the absorbing function $F_{ax} = \exp \left[- 10 \times \left(\frac{x - x_{ab}}{x_0}\right)^2 \right]$ when $x_{ab} < |x| < 55\mathrm{a.u.}$ and $F_{ay} = \exp \left[- 10 \times \left(\frac{y - y_{ab}}{y_0}\right)^2 \right]$ when $y_{ab} < |y| < 55\mathrm{a.u.}$ on the electron wave function at each step to prevent non-physical reflection at the boundary. Where $x_{ab} = y_{ab} = 30\mathrm{a.u.}$ and $x_0 = 55\mathrm{a.u.} - x_{ab}$, $y_0 = 55\mathrm{a.u.} - y_{ab}$ is the length of the absorbing area. The convergence of the results is checked by varying the above parameters. The initial wave function for $\nu = 0$ is obtained by the imaginary propagation method \cite{Bian0}. When we investigate the influence of the higher initial vibrational state, we calculate a 2D-TDSE by using $\psi(x, R, 0) = \chi_{\nu}(R) \psi_0(R, x)$, where $\psi_0(R, x)$ is the ground state of electron at each fixed R and $\chi_{\nu}(R)$ is the nuclear vibrational state with quantum number $\nu$ \cite{Chelkowski1}. We calculate the time-dependent dipole acceleration \cite{Burnett}. Then we Fourier transform it to obtain the HHG spectra:
\begin{equation}\label{E7}
S(\omega) \sim |\hat{\textbf{e}} \cdot \hat{\textbf{a}}(\omega)|^2 = |\int \langle \psi(t)|\hat{\textbf{e}} \cdot \left[\nabla V + \textbf{E} \right]| \psi(t)\rangle e^{ - i \omega t}\mathrm{d}t|^2.
\end{equation}

Eq.(\ref{E7}) gives the spectra in the direction $\hat{\textbf{e}}$, which is chosen to be the direction of the laser field in our investigation.
\section{RESULTS AND DISCUSSION}\label{III}
\begin{figure}
\centering\includegraphics*[width=8 cm]{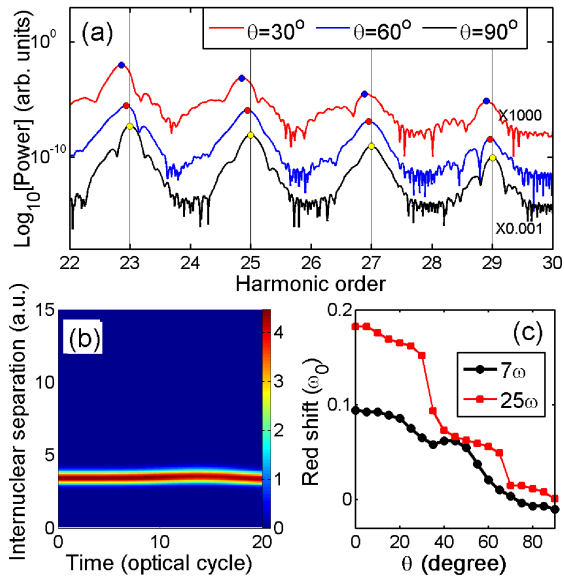}
\caption{(Color online)  (a) HHGs in high energy region when the alignment angle is $30^{\circ}$ (red line), $60^{\circ}$ (blue line) and $90^{\circ}$ (black line) respectively, where the result in $30^{\circ}$ and $90^{\circ}$ are multiplied by 1000 and 0.001 respectively to make them clearly distinguished. The time evolution of the nuclear probability density when $\theta = 90^{\circ}$ is shown in (b). The redshifts of 7th and 25th harmonics as a function of the alignment angle are shown in (c).}
 \label{Fig2}
\end{figure}
\begin{figure}
\centering\includegraphics*[width=8 cm]{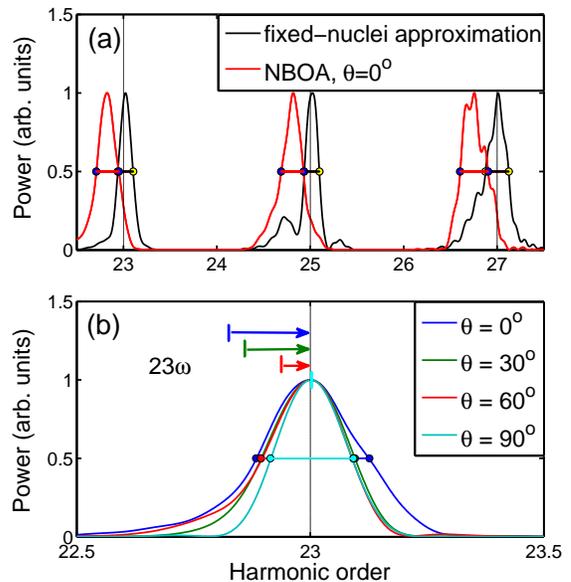}
\caption{(Color online) (a) HHGs in the fixed-nuclei approximation (black line) and NBOA (red line) and their FWHMs when the alignment angle is $0^{\circ}$. (b) The FWHMs of the 23rd order harmonics in different alignment angles in NBOA. The positions of the 23rd order harmonics are moved to be exactly at the $23\omega$ for ease of comparison. The corresponding short vertical lines mark the original positions of the harmonics.}
 \label{Fig3}
\end{figure}

Fig. \ref{Fig1} shows the HHGs from the $\mathrm{H}_2^+$ molecules when they interact with linearly polarized laser pulses at 500 nm, $10^{14}\mathrm{W}/\mathrm{cm}^2$ and the molecules are parallel to the laser polarization direction. Firstly, one can clearly see that redshifts occur considering the nonadiabatic nuclear motion as expected in \cite{Bian, Yuan}. In Fig. \ref{Fig1}(a), harmonic peaks in the fixed-nuclei approximation locate in the odd times of the fundamental frequency exactly, while the photon energies of HHG decrease in NBOA. To illustrate this phenomenon, we calculate the time profile of the HHGs with the wavelet analysis \cite{Antoine, Chandre} in Fig. \ref{Fig1}(b). According to the semiclassical three-step model, during the rising part of the pulse, electrons ionized later will get more energies due to bigger laser amplitude than the previous laser cycle and this results in blueshifts in the HHG spectra. On the contrary, redshifts occur on the falling part \cite{Bian}. When the harmonics generated from the two parts are almost equal, which is the case in the fixed-nuclei approximation in the upper panel of Fig. \ref{Fig1}(b), the blueshifts and the redshifts cancel each other out, and there is no net spectral shift. However, the time profile analysis of the HHG in high energy region in the lower panel of Fig. \ref{Fig1}(b) shows that more harmonics generate on the falling part of the pulse than on the rising part in NBOA, which results in the redshifts dominating. What is the main effect to enhance the efficiency of HHGs on the falling part? We note that the internuclear separation is almost constant at the equivalent distance on the rising part of the pulse, where the ionization rate of the system is negligible. However, the internuclear separation increases dramatically in the second half of the pulse due to the dissociation process as shown in Fig. \ref{Fig1}(c). This causes the ionization potential smaller \cite{Astiaso} and ionization rate bigger \cite{Zuo}. Besides, we find that redshifts in the high energy region (inset at the top-right corner in Fig. \ref{Fig1}(a)) are more obvious than those in the low energy region (inset at the bottom-left corner in Fig. \ref{Fig1}(a)). Fig. \ref{Fig1}(d) also shows the trend that the redshifts become bigger with the increase of the harmonic orders, which is contrary to the trend of redshift due to resonance state \cite{Bian1}. Comparing the wavelet analyses of HHGs between the high and low (not shown) energy region, one can see that higher order harmonics are generated more asymmetrically. This is because that the harmonics in the high energy region are mostly generated after the peak of the pulses and the asymmetry caused by the dissociation is more distinct than the case of strong resonance \cite{Bian1}. This trend may help us distinguish the mechanism of redshift in HHG coming from resonance or dissociation.
\begin{figure}
\centering\includegraphics*[width=8 cm]{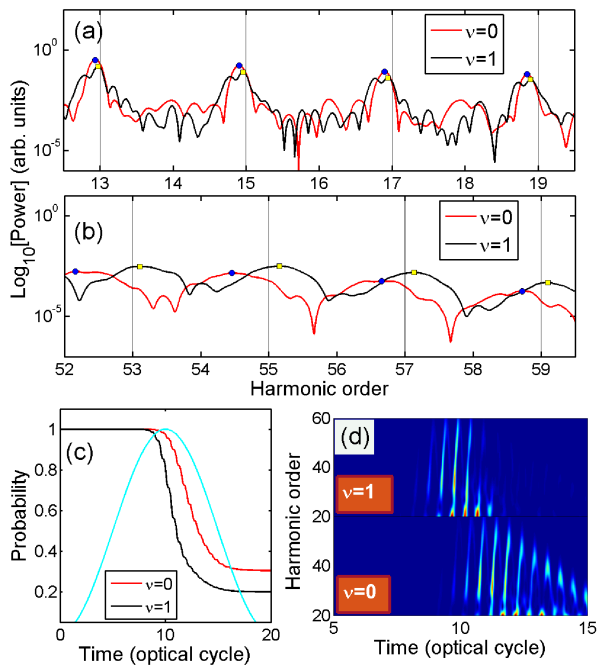}
\caption{(Color online) The HHGs generated from $\nu = 0$ (red line) and $\nu = 1$ (black line) in the low (a) and high (b) energy regions respectively. (c) shows the decreases of probabilities of finding electron in the calculating box in $\nu = 0$ (red line) and $\nu = 1$ (black line). Besides, the laser profile is plotted (cyan line). (d) shows the time profiles of the HHGs in $\nu = 0$ (lower penal) and $\nu = 1$ (upper penal). The laser wavelength is 800nm, and the intensity is $3 \times 10^{14}\mathrm{W}/\mathrm{cm}^2$.}
 \label{Fig4}
\end{figure}

Fig. \ref{Fig2} shows the impact of alignment angles dependence of redshifts. One can see the redshifts decreasing when the alignment angle $\theta$ switches from $30^{\circ}$ to $90^{\circ}$ in Fig. \ref{Fig2}(a). Moreover, the changes of the 7th and 25th order harmonics versus $\theta$ in Fig. \ref{Fig2}(c) also show the inverse proportion between them. The redshifts are most obvious when the molecular axis is parallel to the field polarization direction (Fig. \ref{Fig1}(a)), while there is almost no redshift when they are perpendicular to each other. In fact, the internuclear separation is almost constant when the alignment angle is large, and we present the time evolution of the nuclear probability density when $\theta$ is $90^{\circ}$ in Fig. \ref{Fig2}(b) for example. This is because that the dissociation mainly comes from the dissociating first excited 2$p\sigma_u$ state. When $\theta = 0^{\circ}$, the coupling between the ground state 1$s\sigma_g$ and 2$p\sigma_u$ is the strongest. As $\theta$ increases, the transition probability to 2$p\sigma_u$ state decreases since the electric field along the molecular axis becomes weak. As a result, the influence of the external field on molecular dissociation weakens when the alignment angle increases. Thus the dissociation of $\mathrm{H}_2^+$ is negligible and the redshifts in HHG are not obvious when $\theta = 90^{\circ}$.

Fig. \ref{Fig3} shows the FWHMs of the harmonic peaks in different situations. Comparing with the fixed-nuclei approximation, the peaks are obviously wider under the influence of nuclear motion in NBOA, as shown in Fig. \ref{Fig3}(a), which agrees with the experimental measurements \cite{Yuan}. In addition, the FWHM of certain order harmonics (for example, the 23rd order harmonics in Fig. \ref{Fig3}(b)) decreases with increasing alignment angle, which shows the same trend as the redshifts. Intuitively, the spectrum broadening is caused by the chirp induced by the nuclear motion. And the FWHM naturally become smaller with the enhancement weakening when the alignment angle increases.

We also investigate the dependence of nuclear initial vibrational state on FM in HHG \cite{Zhao}. Fig. \ref{Fig4}(a) and (b) show the FM in HHG spectra of $\mathrm{H}_2^+$ with different initial vibrational states irradiated by pulses at 800 nm, $3 \times 10^{14}\mathrm{W}/\mathrm{cm}^2$ in the low and high energy regions respectively. The redshifts in $\nu = 0$ are obvious in both the high and low energy regions. Nevertheless, the redshifts in $\nu = 1$ are inconspicuous in the low energy region and there are even blueshifts in the high energy region. By using the wavelet analysis in Fig. \ref{Fig4}(d), we can see that when the molecule is in the nuclear initial vibrational state $\nu = 0$, more HHGs generate on the falling part of the pulses due to the delay of dissociation compared to the peak of the laser pulse, while when $\nu = 1$, the dissociation process starts early and the molecules experience a fast depletion. Consequently, the moments of the generation of HHGs in high energy region are mainly advanced to the rising part of the pulse. To figure out the cause in $\nu = 1$, we plot the time-dependent probability of finding electron in the calculation area. As shown in Fig. \ref{Fig4}(c), it is the faster disassociation speed that leads to most electrons freed before the falling part of the pulse and blueshifts occurring.

\section{CONCLUSIONS}\label{IV}

In conclusion, we numerically calculate the 3D-TDSE of $\mathrm{H}_2^+$ interacting with linearly polarized laser pulses in NBOA. The information of molecular alignment angles and initial vibrational states is found to be encoded in the redshifts and the width of the HHG. Redshifts in higher initial vibrational state are weaker because of faster dissociation speed. Increasing alignment angle weakens the impact on disassociation, and then the redshifts become smaller. Nuclear motion also broadens the harmonics peaks in the high energy region. Big redshifts signify that most HHGs generate on the falling part of the laser pulse, which implies the slow reaction of nuclei caused by heavy molecule, low initial vibrational state or small alignment angle. Since Redshifts root in a nonadiabatic response of the molecule and reflect temporally asymmetric emission of HHGs, they can be used to detect the nuclear dynamics and probe electron-nuclei correlation.
\section{ACKNOWLEDGEMENTS}\label{V}

The authors thank Prof. Peng-Fei Lan, Prof. Qing-Bin Zhang, Dr. Tao-Yuan Du and Xin-Qiang Wang very much for helpful discussions. This work is supported by the National Natural Science Foundation of China(Grants No. 11404376 and No. 11561121002).

\end{document}